\documentclass[10pt]{article}
\usepackage{spconf,amsmath,graphicx}
\usepackage{setspace}
\usepackage{amssymb}
\usepackage{graphicx,psfrag} 
\usepackage{epsfig} 
\usepackage{esdiff} 
\usepackage{amsmath,amssymb,mathabx} 
\usepackage{xfrac}   
\usepackage{subfigure}  
\usepackage{pst-sigsys} 
\usepackage{algorithm}
\usepackage{cite}
\usepackage{algorithmic}
\usepackage{colortbl}

\usepackage{setspace}
\usepackage{color}

\newcommand{\beq}{\begin{equation}}
\newcommand{\eeq}{\end{equation}}
\newcommand{\beqa}{\begin{eqnarray}}
\newcommand{\eeqa}{\end{eqnarray}}
\newcommand{\ben}{\begin{enumerate}}
\newcommand{\een}{\end{enumerate}}

\title{Super-Resolution From Binary Measurements With Unknown Threshold}
\name{Subhadip Mukherjee, Anjany Kumar Sekuboyina, and Chandra Sekhar Seelamantula}
\address{Department of Electrical Engineering, Indian Institute of Science, Bangalore 560012, India\\Emails: subhadip@ee.iisc.ernet.in, anjanykumar@outlook.com, chandra.sekhar@ieee.org}
\begin{document}
%
\maketitle

\small
\begin{abstract}
We address the problem of super-resolution of point sources from binary measurements, where random projections of the blurred measurement of the actual signal are encoded using only the sign information. The threshold used for binary quantization is {\it not known} to the decoder.~We develop an algorithm that solves convex programs iteratively and achieves signal recovery. The proposed algorithm, which we refer to as the \textit{binary super-resolution} (BSR) algorithm, recovers point sources with reasonable accuracy, albeit up to a scale factor.~We show through simulations that the BSR algorithm is successful in recovering the locations and the amplitudes of the point sources, even in the presence of significant amount of blurring.~We also propose a framework for handling noisy measurements and demonstrate that BSR gives a reliable reconstruction (correspondingly, reconstruction signal-to-noise ratio (SNR) of about $22$ dB) for a measurement SNR of $15$ dB.
\end{abstract}
\begin{keywords}
Super-resolution, binary compressive sensing, iteratively reweighted $\ell_1$ minimization.  
\end{keywords}

%
\section{Introduction}
Studying the finer details of a signal, beyond the resolution provided by measurement systems, is of importance to many applications such as medical imaging \cite{greenspan}, astronomy \cite{puschmann}, super-resolution localization microscopy \cite{mccutchen, hell, zhuang, physics_today}, etc. The vast body of literature on super-resolution deals with the problem of recovering fine details of a signal from coarser measurements. 
In most super-resolution techniques, one essentially extrapolates the spectra of the signals of interest to higher frequencies, given their low-frequency measurements. As a concrete example, consider astronomical imaging, where one is interested in identifying and super-resolving celestial objects that appear as point sources in the images (often blurred) captured by telescopes. For several applications, it would be desirable to identify close-by sources, surpassing the barrier posed by the diffraction limit of the imaging device.\\ 
\indent Super-resolution of point sources has been recently studied by Cand\`es and Granda, for noiseless \cite{candes_noiseless_SR} and noisy \cite{candes_noisy_SR} measurements. They addressed the problem of extrapolating the spectrum of a superposition of point sources up to a frequency $f_{\text{hi}}$, from its low-pass measurements up to a frequency $f_{\text{lo}}$, and showed that a convex optimization-based approach produces a stable estimate, such that the reconstruction error is proportional to the noise level times the square of $\frac{f_{\text{hi}}}{f_{\text{lo}}}$, provided the sources are separated by at least $\frac{2}{f_{\text{lo}}}$.\\
\indent An important aspect that one needs to consider while building an acquisition system is the precision of measurement, dictating the cost, speed of acquisition, and the subsequent storage requirement. From a practical viewpoint, it is important to design a signal recovery algorithm that takes into account the effect of measurement quantization. It is a worthwhile exercise to extend the limits of the reconstruction techniques where the signal is measured with finite precision. Designing compressive imaging systems with one-bit measurements was considered by Bourquard et al. \cite{binary_imaging1, binary_imaging2}, who proposed an acquisition hardware and reconstruction algorithm for image recovery starting from binary measurements.~Our work focuses on developing an algorithm for the separation of point sources in the extreme case where the imaging device employs one-bit quantization in comparison against an {\it unknown threshold} and records only binary data, potentially leading to fast and inexpensive hardware. Our reconstruction approach is iterative and based on convex programming, yielding accurate estimates up to a scale factor.\\
\indent Sparse recovery using binary measurements was first addressed by Boufounous and Baraniuk \cite{baraniuk1}, who proposed a fixed-point continuation algorithm, treating the measurements as sign constraints and restricting the search space to the unit sphere to avoid degenerate solution.~Gupta et al. \cite{nowak1} proposed an adaptive algorithm for support recovery from sign measurements, requiring a sample complexity of $\mathcal{O}\left(s \log l_x\right)$, where $s$ and $l_x$ denote the sparsity and signal dimension, respectively.~Plan and Vershynin developed a linear programming-based sparse recovery algorithm from binary measurements \cite{plan1}, extendable to approximately sparse vectors. They also established a connection between one-bit compressive sensing (CS) and sparse logistic regression and developed a convex program-based approach robust to sign-flips \cite{plan2}. Aside from the convex-relaxation-based approaches, greedy algorithms such as matching sign pursuit \cite{msp}, adaptive outlier pursuit \cite{aop} and iterative algorithms such as binary iterative hard thresholding \cite{biht}, randomized fixed-point iterations \cite{rfpi}, etc. are also available in the literature. The suitability of the one-bit CS paradigm over its multibit counterpart, for several practical applications, and its robustness to non-linear distortions were studied in \cite{onebit_multibit} and \cite{boufounous1234}, respectively.\\
\indent The threshold used for the one-bit quantizer in the context of CS plays a crucial role in the reconstruction performance.~Adaptive algorithms for selecting the threshold have been developed by Kamilov et al. \cite{kamilov} and Fang et al. \cite{fang}. In the following, we highlight our contributions and indicate the major differences between our approach and the related algorithms.

\subsection{Our Contribution}
\indent We address the problem of recovering a combination of point sources using a setup that acquires binary measurements, where only the sign information of the random linear projections of a blurred measurement of the actual source is recorded. The threshold used for collecting the binary measurements is not known to the decoder/reconstruction algorithm unlike the frameworks considered in \cite{kamilov, fang}. In fact, our reconstruction algorithm also gives an estimate of the threshold jointly with the underlying signal starting from the binary measurements. We formulate the problem of signal recovery, albeit up to a scale factor, as a convex optimization, and develop a \textit{binary super-resolution} (BSR) algorithm, leveraging the idea of \textit{iteratively reweighted $\ell_1$ minimization} (IR$\ell_1$) \cite{reweighted_L1}. We demonstrate successful application of the BSR algorithm in recovering point sources and demonstrate its robustness to noise.
\section{Problem Statement and The BSR Algorithm}
\label{sec_problem_form}
Consider a sparse signal $x(n)=\sum_{j=1}^{s}\alpha_j \delta \left(n-n_j\right)$, consisting of $s$ Kronecker impulses of amplitudes $\alpha_j$, located at time instants $n_j$. Suppose $x(n)$ is acquired through a measurement device having a low-pass impulse response $h(n)$, giving rise to the measured signal $z(n) = \sum_{j=1}^{s}\alpha_j h \left(n-n_j\right)$. Instead of directly recording $z(n)$, we consider a measurement setup where one computes linear projections of $z(n)$ using random sensing signals $p_i\left(n\right)$, $1\leq i \leq m$, compares them with a threshold $\tau$, and encodes them as $+1$ or $-1$, depending on whether they exceed $\tau$ or not, respectively. More precisely, the measurement device collects binary measurements of the form $y_i = \text{sgn}\left(\left \langle z(n), p_i(n) \right \rangle -\tau\right), 1\leq i \leq m$, where $\langle \cdot,\cdot \rangle$ indicates the inner-product, and sgn is the signum function. The encoder chooses a value of $\tau$ such that not all measurements are mapped to $+1$ or $-1$. In fact, the encoder could make an optimal choice of $\tau$ according to some predefined criterion. The measurement process is expressed concisely as 
\begin{equation}
y_i = \text{sgn}\left( A\bold z -\tau \right)_i=\text{sgn}\left( AH \bold x -\tau \right)_i,
\label{measurement_eq1}
\end{equation}
where $\bold x$ denotes the vector representation of $x(n)$, $H$ denotes the linear convolution matrix corresponding to the impulse response $h(n)$, and the $i^{\text{th}}$ row of $A$ contains the random sensing signal $p_i\left(n\right)$.~Given binary measurements as in \eqref{measurement_eq1}, our objective is to recover the underlying signal $\bold x$, or, equivalently, to estimate the amplitudes and the locations of the Kronecker impulses. 
\subsection{Development of the BSR Algorithm}
\label{algo_development}
Defining $\Phi=AH$, the binary measurements in \eqref{measurement_eq1} are written as $y_i = \text{sgn}\left( \phi_i^T \bold x -\tau \right)$, where $\phi_i$ denotes the $i^{\text{th}}$ row of $\Phi$. We seek to recover a sparse vector $\bold x$, with the least number of non-zero entries, together with an appropriate $\tau$, such that the recovered $\left(\hat{\bold x},\hat{\tau}\right)$ pair is consistent with the measurements. This is accomplished by solving
\begin{equation}
\left(\hat{\bold x},\hat{\tau}\right)=\arg \underset{\bold x, \tau}{\min}\text{\,\,}\left\| \bold x \right\|_0  \text{\,subject to\,\,}
y_i \left(\phi_i^T \bold x -\tau\right)\geq 0, 1\leq i \leq m.
\label{noiseless_SVM_onebit}
\end{equation}
The inequality constraints in \eqref{noiseless_SVM_onebit} ensure that the recovery is consistent with the sign measurements. Although the optimization posed in (\ref{noiseless_SVM_onebit}) captures the objective of sparse super-resolution from sign measurements, there are two major defects associated with it, namely, non-convexity and degeneracy. Non-convexity arises because of the $\ell_0$-norm objective, whereas degeneracy is a consequence of the fact that $\hat{\bold x}=0$, $\hat{\tau}=0$ are feasible and yield the minimum possible value of the objective. However, it is possible to eliminate either defects as explained in the following. 
\subsubsection{Eliminating degeneracy and non-convexity}
\label{degeneracy_removal}
Since $y_i =\text{sgn}\left(\phi_i^T \bold x -\tau\right)$, the measurement consistency requirements can be expressed as 
\begin{equation}  
y_i \left(\phi_i^T \bold x -\tau\right)> \delta, 1\leq i \leq m,
\label{linear_separability}
\end{equation}
where $\delta = \underset{1\leq i \leq m}{\min}\text{\,}y_i \left(\phi_i^T \bold x -\tau\right)\geq 0$. If the underlying signal $\bold x$ follows a continuous distribution, we have that $\delta>0$, with probability one.~Dividing both sides of \eqref{linear_separability} by $\delta$ and absorbing $\delta$ in $\bold x$ and $\tau$, the inequalities in (\ref{linear_separability}) can be rewritten in an equivalent form, given by $y_i\left(\phi_i^T \bold x- \tau\right) \geq 1$, for all $i$. Thus, one can avoid degenerate solution by rewriting (\ref{noiseless_SVM_onebit}) as
\begin{equation}
\left(\hat{\bold x},\hat{\tau}\right)=\arg \underset{\bold x, \tau}{\min}\text{\,\,}\left\| \bold x \right\|_0  \text{\,\,subject to\,\,}
y_i \left(\phi_i^T \bold x -\tau\right)\geq 1.
\label{noiseless_SVM_onebit_no_degeneracy}
\end{equation}
Although the solution to \eqref{noiseless_SVM_onebit_no_degeneracy} is not degenerate, it suffers from scale ambiguity.~To illustrate further, consider a solution $\left(\hat{\bold x},\hat{\tau}\right)$ to \eqref{noiseless_SVM_onebit_no_degeneracy}. For any $\beta>1$, $\left(\beta\hat{\bold x},\beta\hat{\tau}\right)$ is also a solution, since it is feasible and yields the same value of the objective. To avoid this ambiguity, we assume that $\left\| \bold x \right\|_2=1$, and this condition is enforced after computing a solution to \eqref{noiseless_SVM_onebit_no_degeneracy}.\\
\indent In order to avoid non-convex optimization, we employ IR$\ell_1$ to approximate the solution to (\ref{noiseless_SVM_onebit_no_degeneracy}). In the $p^{\text{th}}$ iteration, we solve
\begin{eqnarray}
\left(\hat{\bold x}^{(p)},\hat{\tau}^{(p)}\right)=\arg \underset{\bold x, \tau}{\min}\text{\,\,}\left\| \Lambda^{(p)} \bold x \right\|_1  \text{\,\,s.t.\,\,}y_i \left(\phi_i^T \bold x -\tau\right)\geq 1,
\label{noiseless_SVM_multibit_nondegen_cvx}
\end{eqnarray} 
where $ \Lambda^{(p)}$ is a diagonal weight matrix, initialized with $\lambda^{(0)}_i=1$, $\forall i$, and modified as $\lambda^{(p)}_i=\frac{1}{\left| \hat{\bold x}_i^{(p-1)} \right|+\epsilon}$, for every $p$. A constant $\epsilon>0$ is added to the denominator for numerical stability. It is shown in \cite{reweighted_L1} that the IR$\ell_1$ algorithm arises out of minimizing an objective of the form $f_0\left(\bold x\right)=\sum_{i=1}^{l_x}\log \left(\left| x_i \right|+\epsilon \right)$. The iterations are repeated until $p = I$, for some predetermined $I$. A convex optimization software, namely CVX \cite{grant}, is employed to solve \eqref{noiseless_SVM_multibit_nondegen_cvx}. The BSR procedure is summarized in Algorithm 1. 

\begin{algorithm}[t]

	\caption{\small{BSR algorithm to estimate a superposition of point sources $\bold x$ from binary measurements $\bold y$ as in \eqref{measurement_eq1}}.}
	\small
	\begin{algorithmic}
		\STATE {\bf 1.} {\bf Input:} The matrix $\Phi=AH$ and the iteration count $I$. 
		\vspace{0.02in}
		\STATE {\bf  2.} {\bf Initialization:} Set $p\leftarrow 0$, $\lambda^{(p)}_i=1$, $1\leq i \leq l_x$, where $l_x$ denotes the length of $\bold x$.
				
		\vspace{0.02in}
		\STATE {\bf  3.} {\bf Iterate the following steps $I$ times:} 
		\begin{enumerate}
			\item Obtain $\left(\hat{\bold x}^{(p)},\hat{\tau}^{(p)}\right)$ by solving \eqref{noiseless_SVM_multibit_nondegen_cvx}.

			\item Update $\lambda^{(p)}_i=\frac{1}{\left| \hat{\bold x}_i^{(p-1)} \right|+\epsilon}$, for all $i$, and set $p \leftarrow p+1$.
		\end{enumerate}
		\STATE {\bf  4.} {\bf Scaling:} Compute $\hat{\bold x} = \frac{\hat{\bold x}^{(I)}}{\left\| \hat{\bold x}^{(I)} \right\|_2}$.
		\STATE {\bf  5.} {\bf Output:} Estimated signal $\hat{\bold x}$.
	\end{algorithmic}
\end{algorithm}

\subsection{Illustration Using One Dimensional (1-D) Signals}
\begin{center}
\begin{figure}[t]
\centering
		\subfigure[\scriptsize{Ground truth $x(n)$}]{
		\includegraphics[width=1.05in]{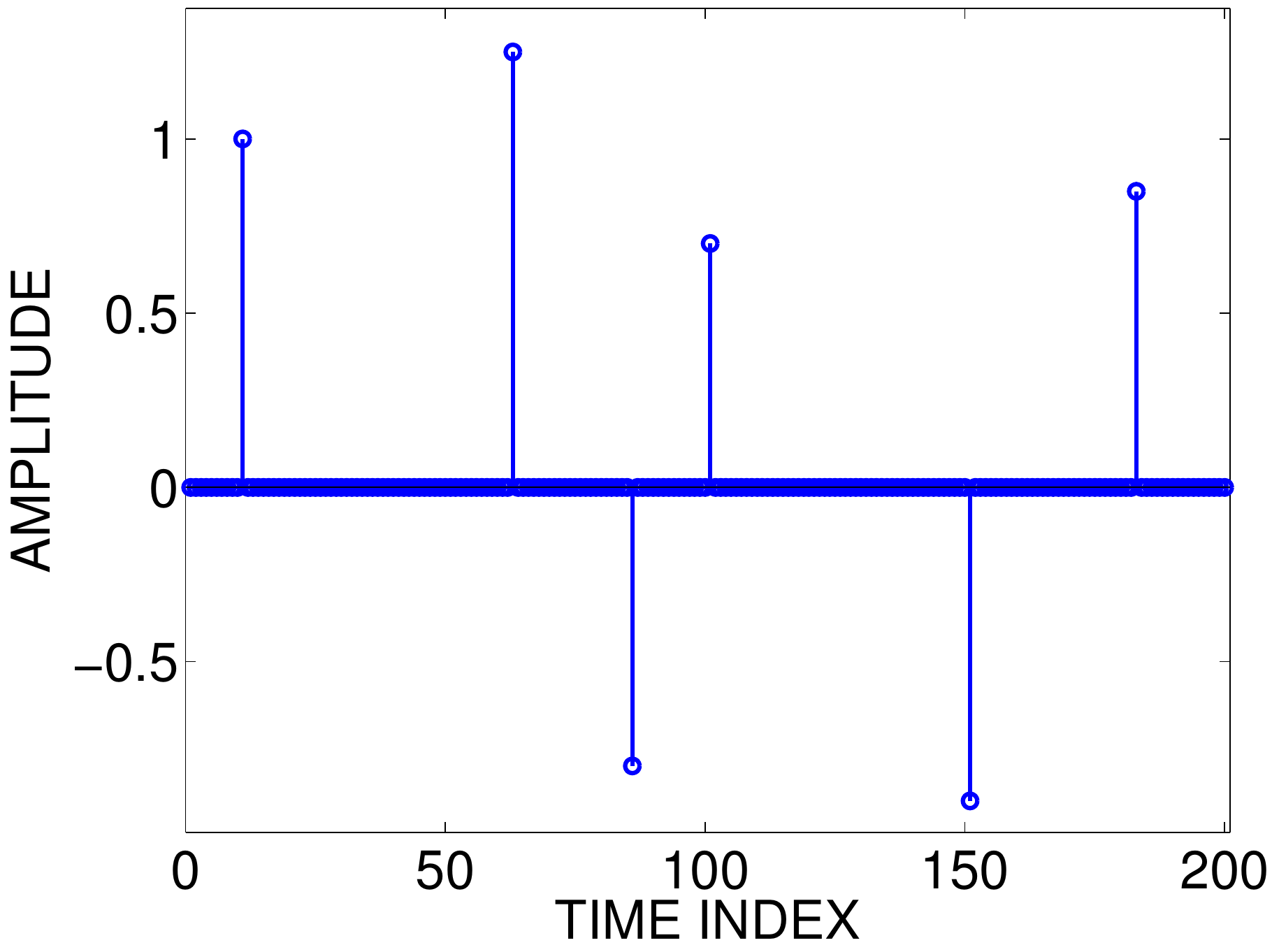}}
		\subfigure[\scriptsize{Filter response $h(n)$}]{
		\includegraphics[width=1.05in]{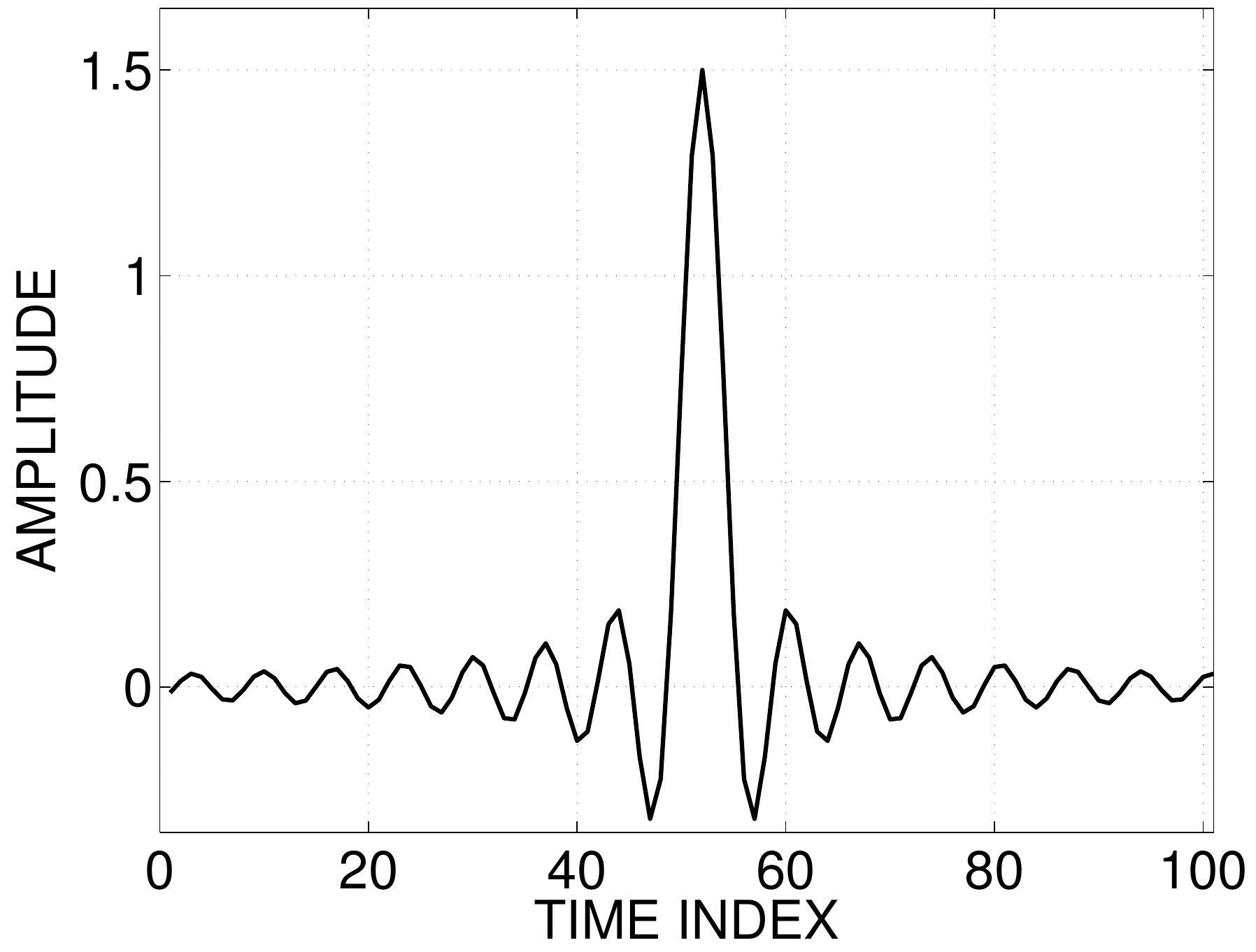}}
		\subfigure[\scriptsize{$z(n)=x(n)\star h(n)$} ]{
		\includegraphics[width=1.05in]{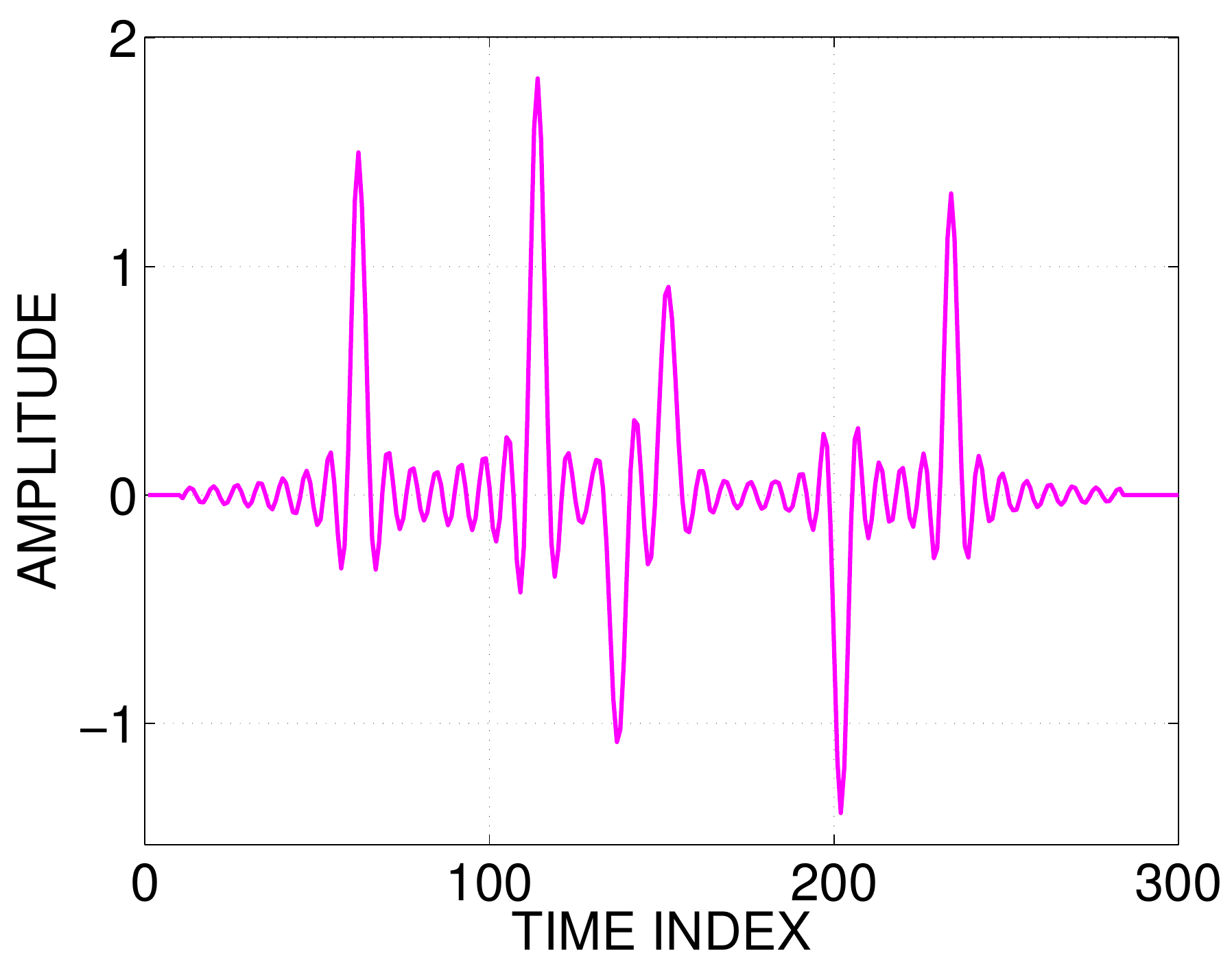}}\\
		\subfigure[\scriptsize{Binary Measurements}]{
		\includegraphics[width=1.05in]{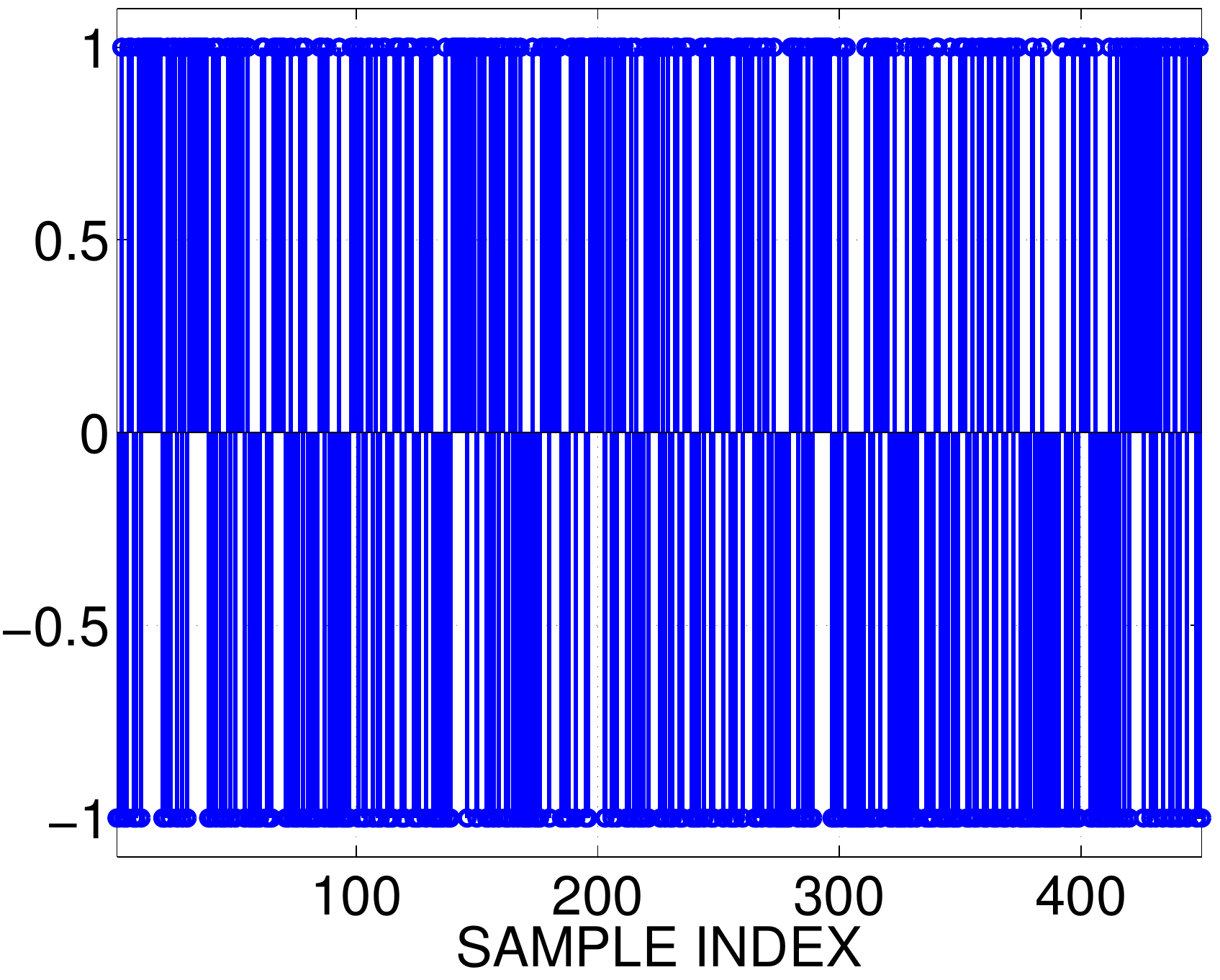}}
		\subfigure[\scriptsize{Reconstructed signal}]{
		\includegraphics[width=1.05in]{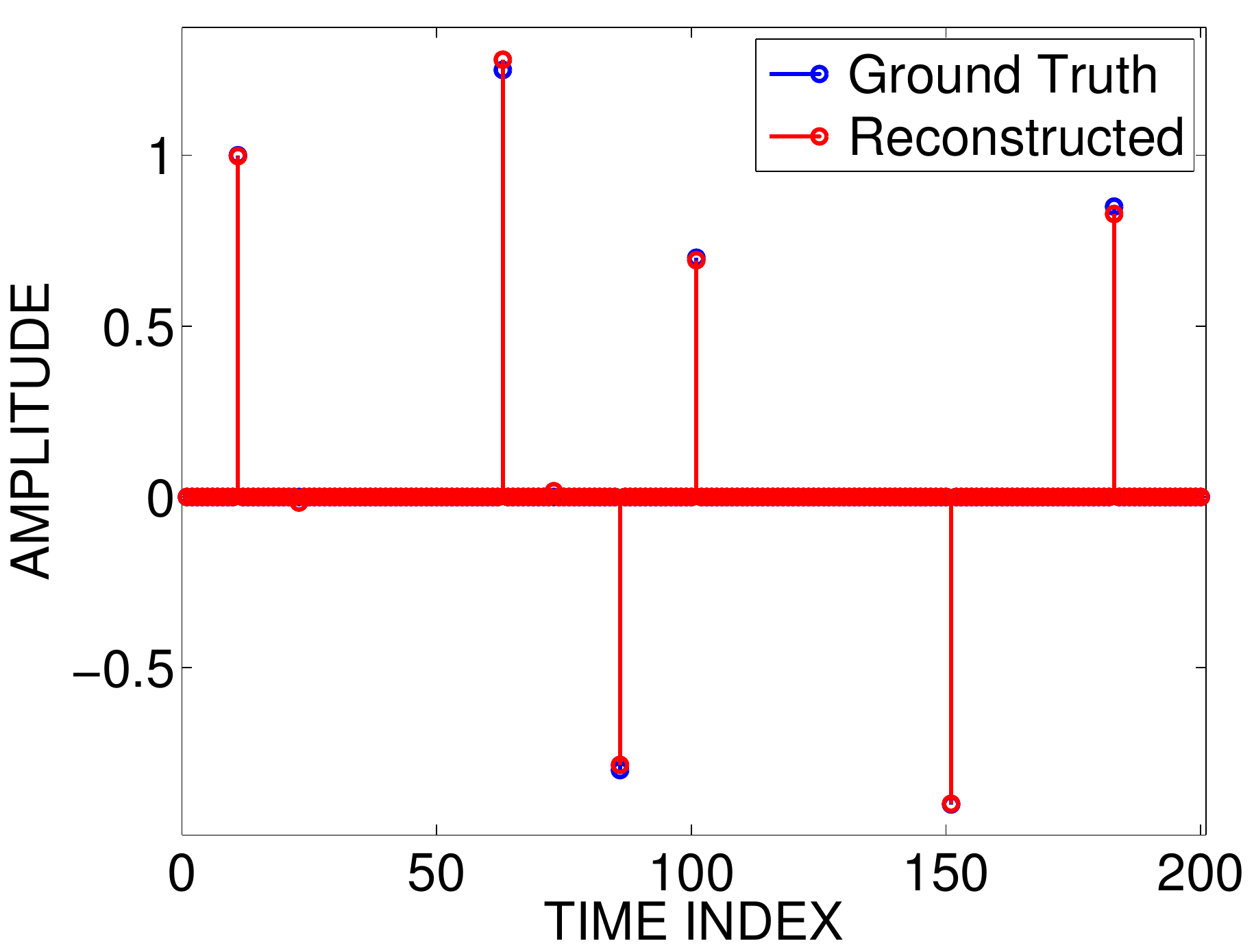}}
		\subfigure[\scriptsize{Recovery SNR}]{
		\includegraphics[width=1.05in]{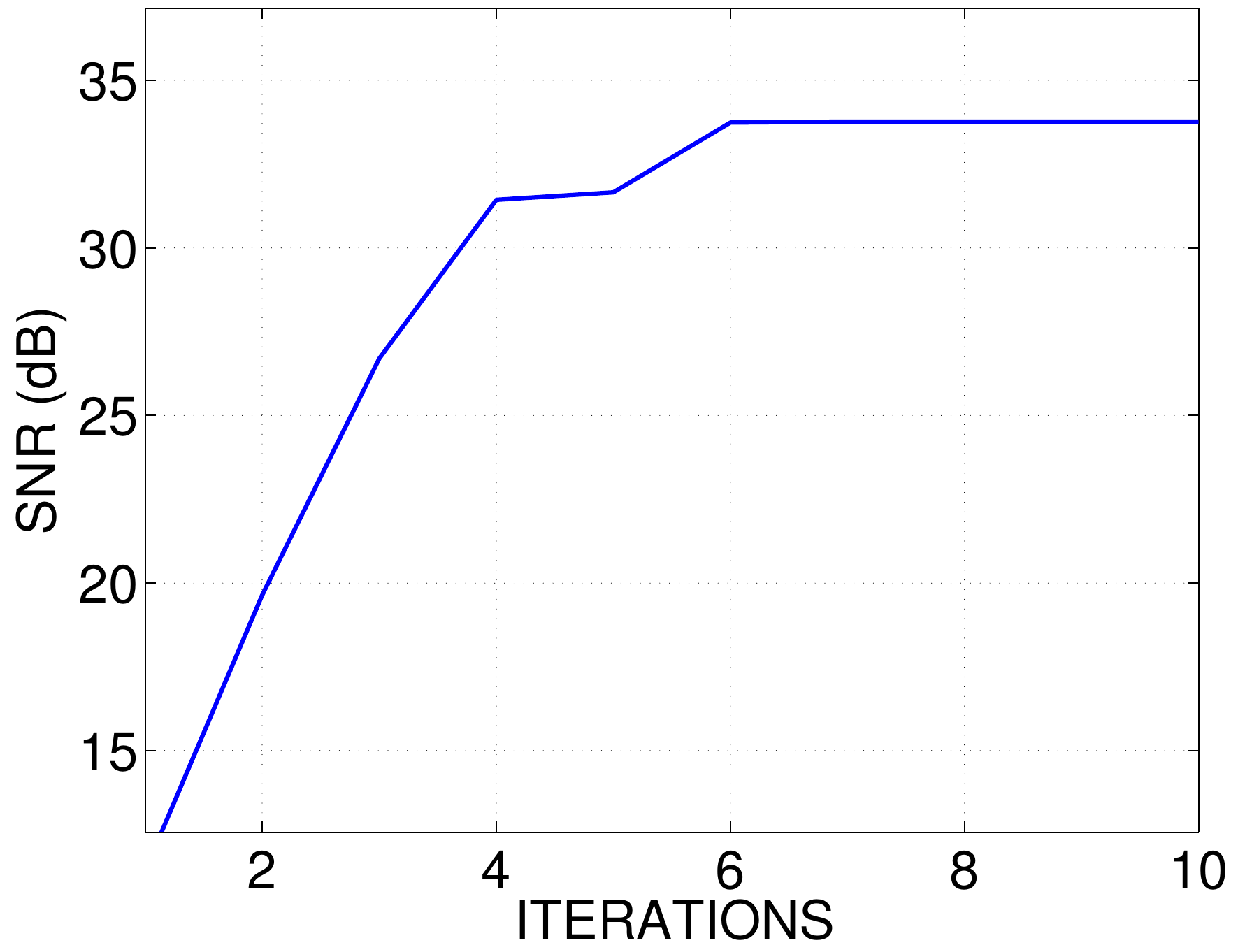}}

	\caption{\small{(Color Online) Performance of BSR on 1-D signals.}}
\label{OneD_illustration_figure}
\end{figure}
\end{center}

To demonstrate the performance of the BSR algorithm, we consider a signal $x(n)$ of length $l_x=200$, containing six impulses, as shown in Fig.~\ref{OneD_illustration_figure}(a). The signal is measured using a device having a low-pass response $h(n)$, which is assumed to be a $\text{sinc}$ function, shown in Fig.~\ref{OneD_illustration_figure}(b). The low-pass measurement $z(n)$ of length $l_z=300$ and the binary measurements $y_i$, $1\leq i \leq m$, are plotted in Figures \ref{OneD_illustration_figure}(c) and \ref{OneD_illustration_figure}(d), respectively. We acquire $m=1.5l_z=450$ binary measurements of the form given in \eqref{measurement_eq1}, where $\tau=-0.1$ and the sensing signals $p_i(n)$ are drawn independently, such that each entry of them takes a value of $+1$ or $-1$ with equal probability. The reconstructed signal, obtained using the BSR algorithm, and the variation of the signal-to-noise ratio (SNR), as a function of iterations, are shown in Figures \ref{OneD_illustration_figure}(e) and \ref{OneD_illustration_figure}(f), respectively. Using the measurement model \eqref{measurement_eq1}, one requires to spend only $b=1.5l_z$ bits for encoding the measurements, whereas, directly encoding $z(n)$ with a precision of $16$ bits for every entry requires $b=16l_z$ bits of storage. Therefore, the BSR algorithm results in a significant reduction in storage, without compromising on the reconstruction accuracy. The idea of using IR$\ell_1$ improves the SNR as the iterations progress, and the reconstructed signal after $10$ iterations matches closely with the ground truth, as can be seen in Fig. \ref{OneD_illustration_figure}(e). 
\vspace{-0.3in}
\section{Experimental Results}
We carry out experiments on images containing point sources to assess the efficacy of the BSR algorithm. A patch-based implementation of the BSR algorithm is developed in order to achieve scalability for images of larger sizes. We also conduct an empirical study to evaluate the performance of BSR under noise. 
\subsection{Patch-Based Implementation of BSR for Images}
Let $\bold x\in \mathbb{R}^{l_x\times l_x}$ and $\bold z\in \mathbb{R}^{l_z\times l_z}$ denote the actual and the blurred images, where $\bold z$ is obtained by applying a low-pass mask $h$ of size $p\times p$ on $\bold x$. For convenience, we assume that $p$ is odd. Consider a patch of size $d\times d$, located at $(i,j)$ in the blurred image $\bold z$. Using the vectorized representation $z^{(i,j)} \in \mathbb{R}^{d^2\times 1}$ of the patch, we write
\begin{equation}
z^{(i,j)} = \mathcal{H}x^{(i,j)},
\label{mask_operation_vec}
\end{equation}
where $x^{(i,j)}\in \mathbb{R}^{\left(d+p-1\right)^2\times 1}$ denotes the vectorized version of the  patch $x\left(i-\frac{p-1}{2}:i+d-1+\frac{p-1}{2},j-\frac{p-1}{2}:j+d-1+\frac{p-1}{2}\right)$ of size $\left(d+p-1\right)\times \left(d+p-1\right)$ in the actual image $x$, and $\mathcal{H}\in \mathbb{R}^{d^2 \times \left(d+p-1\right)^2}$ is a matrix that captures the effect of the mask $h$ on $x$. Binary measurements of the form
\begin{equation}
y_k^{(i,j)} = \text{sgn}\left(a_k^T \mathcal{H}x^{(i,j)}-\tau\right), 1\leq k \leq m,
\label{measurement_eq}
\end{equation}
are acquired for each patch ${(i,j)}$ in $\bold z$ without overlap, where $a_k$ denotes a random vector of size $d^2$, for every $k$. The number of measurements for every patch is chosen as $m=2d^2$. The threshold $\tau$ is unknown to the BSR algorithm and is estimated from the measurements. An estimate of the ground truth $\bold x$ is formed by stitching the recovered patches $\hat{x}^{(i,j)}$, considering the overlap of $(p-1)$ pixels. The number of iterations in the BSR algorithm is taken as $I=5$. 
\subsection{Performance Evaluation on Point-Source Images}
\indent Images of size $256\times 256$, consisting of a superposition of $s=100$ point sources are considered for experimental validation. The locations of the sources are selected uniformly at random from $256^2$ possible locations and the magnitudes are drawn from a uniform distribution over $[0,5]$. Gaussian blur kernels of different sizes $p$ and standard deviations $\sigma_h$ are applied on the ground truth image to obtain a low-pass measurement. Finally, random projections are computed for non-overlapping patches of size $16\times 16$ extracted from the blurred image, followed by one-bit quantization. The number of binary measurements recorded for each patch is $m=2\times 16^2=512$. To reconstruct the ground truth image, the BSR algorithm is iterated $I=5$ times. The performance is measured in terms of the ability to identify the locations and the amplitudes of the point sources.\\
\indent Although the knowledge of the number of impulses $s$ is not assumed in the BSR algorithm, we retain $s$ dominant impulses in the recovered image for the purpose of comparison with the ground truth. To measure the accuracy of BSR in recovering the locations of the impulses, we evaluate the {true positive ratio} (TPR), defined as  $\text{TPR}=\frac{\left|S \cap \hat{S}\right|}{\left|S \right|}$, where $S$ and $\hat{S}$ denote the sets of locations of the impulses in the ground truth and the estimated images, respectively. The set $\hat{S}$ is computed after picking the $s$ most significant impulses in the recovered image, ignoring the ones with negligibly small amplitudes. The TPR metric indicates the fraction of impulses whose locations are identified correctly. The amplitude recovery performance is measured by calculating the SNR over the correctly estimated impulses, denoted as $\text{SNR}_1$, and defined by $\text{SNR}_1=\frac{\left\| \bold x_{S \cap \hat{S}} \right\|_F^2}{\left\| \bold x_{S \cap \hat{S}}-\hat{\bold x}_{S \cap \hat{S}} \right\|_F^2}$, capturing the discrepancy in amplitude estimation over the rightly identified impulses. We compute another metric referred to as the \textit{relative error} (RE), defined as $\text{RE}=\frac{\left\|\hat{\bold x} - \hat{\bold x}_{\hat{S}}\right\|_F^2}{\left\|\bold x\right\|_F^2}$, where $\hat{\bold x}_{\hat{S}}$ indicates an image whose entries are same as $\hat{\bold x}$ over $\hat{S}$ and zero elsewhere. RE measures the sum of the squared amplitudes of the spuriously recovered impulses relative to the energy of the ground truth. Lower values of the RE metric indicate that the spurious impulses have smaller amplitudes, thereby guaranteeing reliable reconstruction.   
\begin{table}[t]
\centering\begin{tabular}{|c| c| c| c|}
\hline 
Kernel parameters  & TPR  & $\text{SNR}_1$ (dB)& RE (dB) \\ 
\hline  $p=5$, $\sigma_h=2$  & $1.0$ &  $28.40$ & $-26.57$  \\ 
  $p=7$, $\sigma_h=3$   & $1.0$  &   $25.09$ & $-17.92$ \\  
$p=9$, $\sigma_h=4$   & $1.0$  &    $22.26$ & $-15.64$ \\  
$p=11$, $\sigma_h=5$   & $1.0$  &   $18.99$ & $-15.13$   \\  
$p=13$, $\sigma_h=6$   & $1.0$  &  $16.47$ & $-16.86$ \\  
$p=15$, $\sigma_h=7$   & $1.0$  &   $13.04$  & $-14.86$ \\  
\hline 
\end{tabular}
\caption{\small{Performance evaluation of the BSR algorithm in estimating images containing point sources; the Gaussian kernel is of size $p \times p$ pixels and has a standard deviation of $\sigma_h$.\vspace{-0.01in}}}
\label{table_performance} 
\end{table}

\begin{figure}[t]
\centering
		\subfigure[\scriptsize{Ground truth}]{
		\includegraphics[width=1.5in]{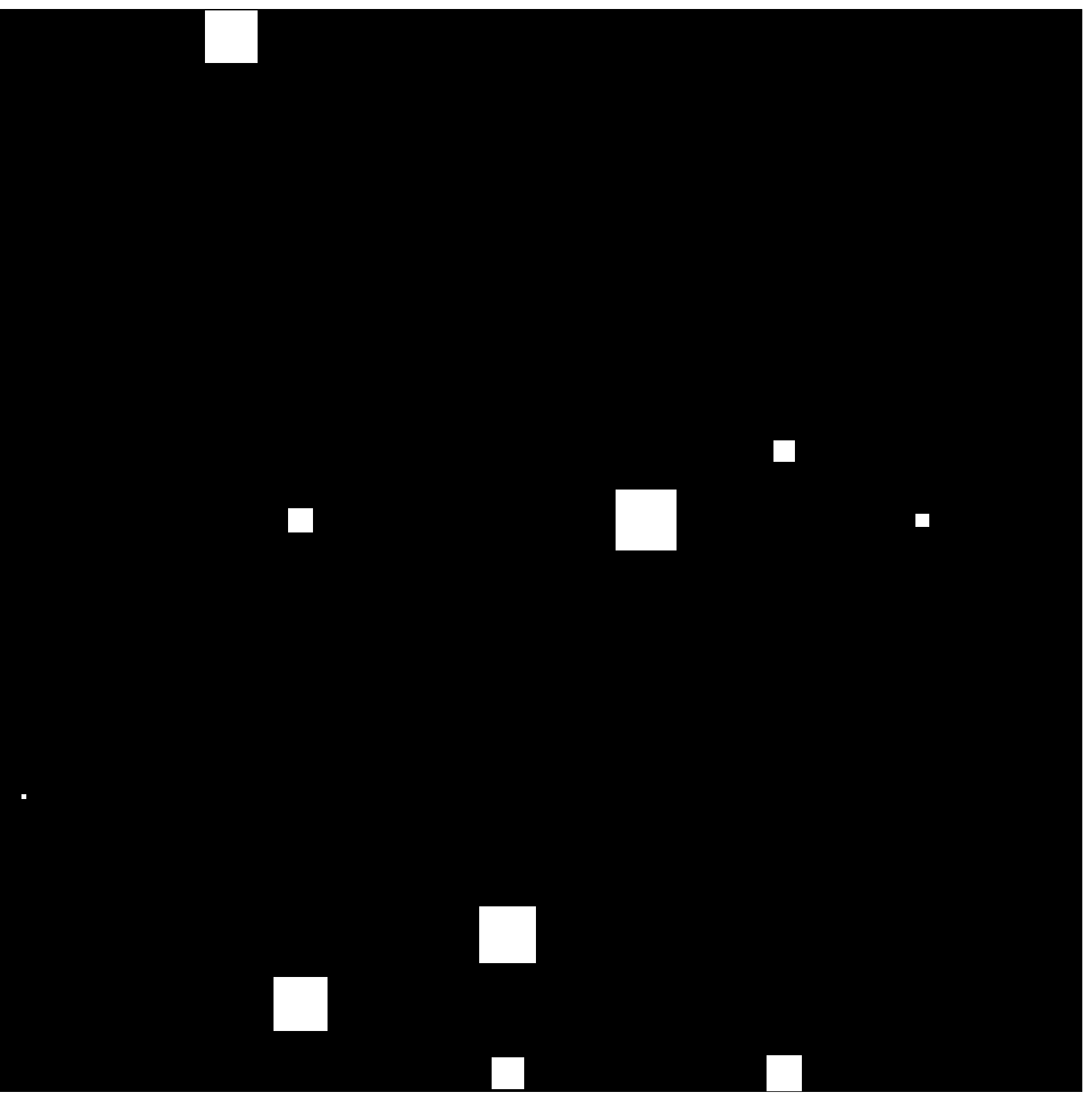}}
		\subfigure[\scriptsize{Blurred measurement}]{
		\includegraphics[width=1.5in]{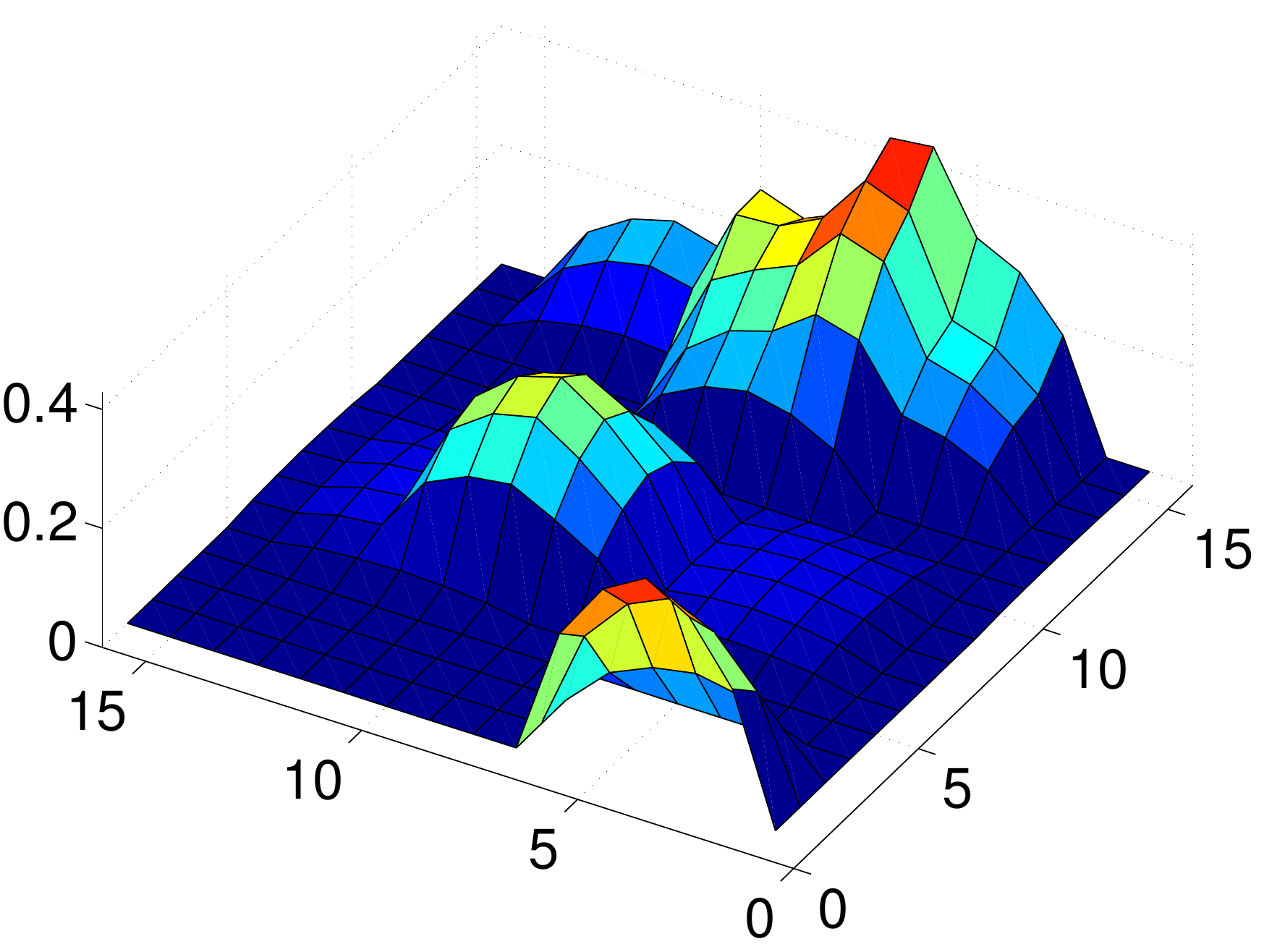}}\\
		\subfigure[\scriptsize{Oracle recovery}]{
		\includegraphics[width=1.5in]{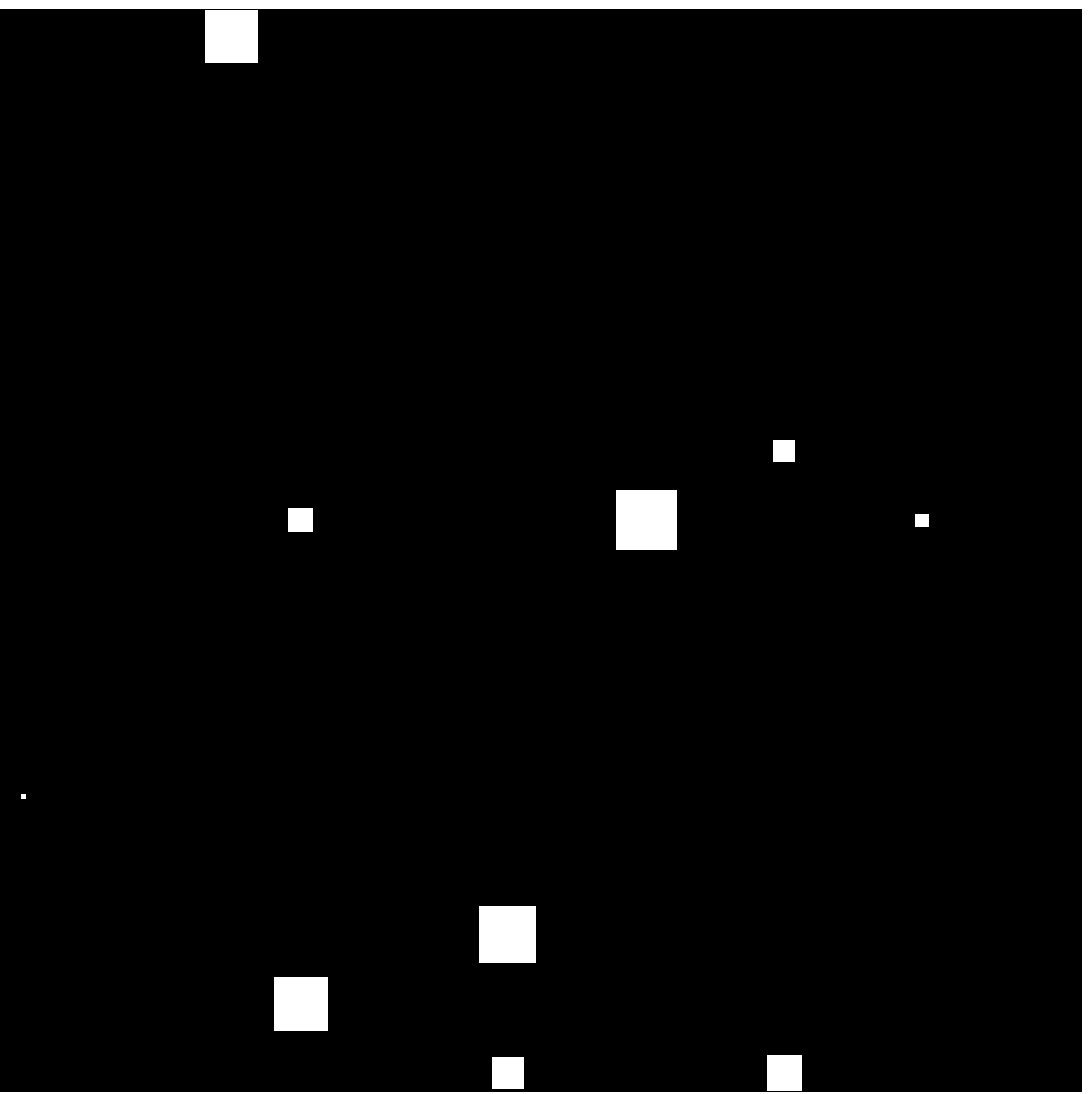}}
		\subfigure[\scriptsize{BSR recovery}]{
		\includegraphics[width=1.5in]{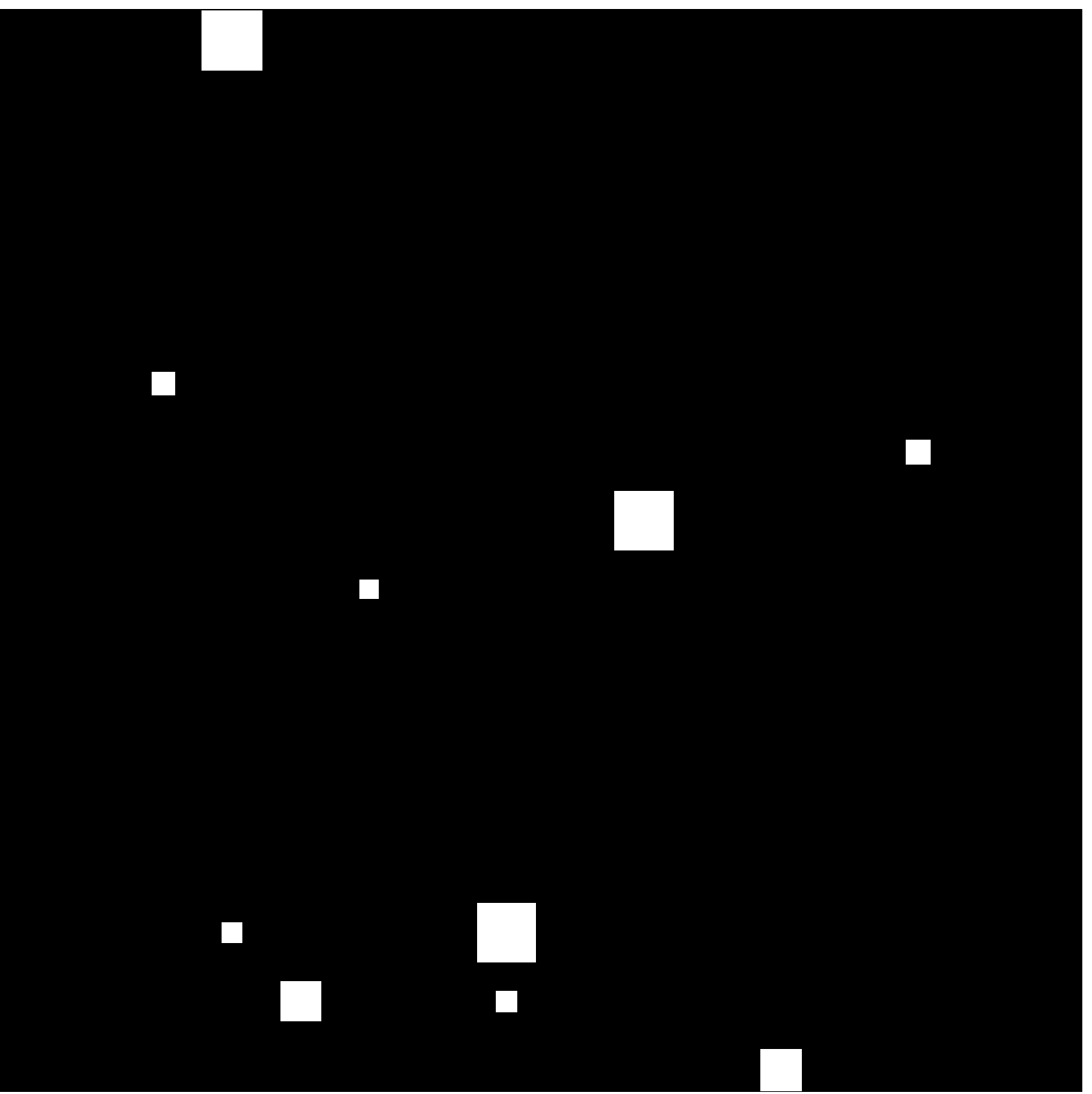}}
			\caption{\small{(Color Online) Performance of BSR on a point-source image, consisting of $s=10$ impulses, shown using Hinton diagrams. The locations and the amplitudes of the impulses are accurately recovered. The kernel parameters are chosen as $p=5$ and $\sigma_h=2$.\vspace{-0.15in}}}
\label{image_figure}
\end{figure}
\indent The values of TPR, $\text{SNR}_1$, and RE obtained using the BSR algorithm are reported in Table \ref{table_performance}, for different values of $p$ and $\sigma_h$. We observe that the BSR algorithm accurately estimates the locations of the point sources, resulting in $\text{TPR}=1.0$. The values of $\text{SNR}_1$ and RE are acceptable for most applications, even for a relatively larger spread and size of the Gaussian kernel. However, increasing the values of $p$ and $\sigma_h$ intensifies the blurring effect, thereby resulting in smaller values of $\text{SNR}_1$ and larger values of RE, as one would typically expect. Despite the deterioration in the performance in terms of $\text{SNR}_1$ and RE, with increase in $p$ and $\sigma_h$, we observe that the BSR algorithm provides a reasonably good estimate of the locations of the impulses in the ground truth.\\    
\noindent For visual comparison, we represent the ground truth and the images reconstructed using the BSR algorithm for a Gaussian blur kernel with $p=5$ and $\sigma_h=2$, in Fig. \ref{image_figure}, using a Hinton diagram \cite{hinton} that is quite popular in the \textit{deep learning} community. The sizes of the squares in the diagram are proportional to the amplitudes of the respective impulses. The ground truth in the experiment consists of $10$ point sources. We compare BSR with a deconvolution setup that acquires measurements without quantization, and the ground truth is recovered by solving 
\begin{equation}
\hat{\bold x}_{\text{oracle}}=\arg \underset{\bold x}{\min}\text{\,\,}\left\| \bold x \right\|_1  \text{\,\,subject to\,\,}\bold y_{\text{fp}} = AH\bold x,
\label{BP_opt}
\end{equation}
where $\bold y_{\text{fp}} $ denotes the full-precision measurement.~The recovered signal using \eqref{BP_opt} is shown in Fig.~\ref{image_figure}(c) and indicated as the \textit{oracle} recovery. We observe in Fig.~\ref{image_figure}(d) that the BSR algorithm recovers the impulses with almost the same accuracy as the oracle, despite being supplied with binary measurements. This result is encouraging, as one can potentially leverage it to build fast and inexpensive systems for resolving point sources in a real-world imaging application.       
\vspace{-0.1in}
\subsection{Performance Evaluation in Noise}
\begin{figure}[t] 
\centering

		\subfigure[\scriptsize{Gaussian blur kernel}]{
		\includegraphics[width=1.5in]{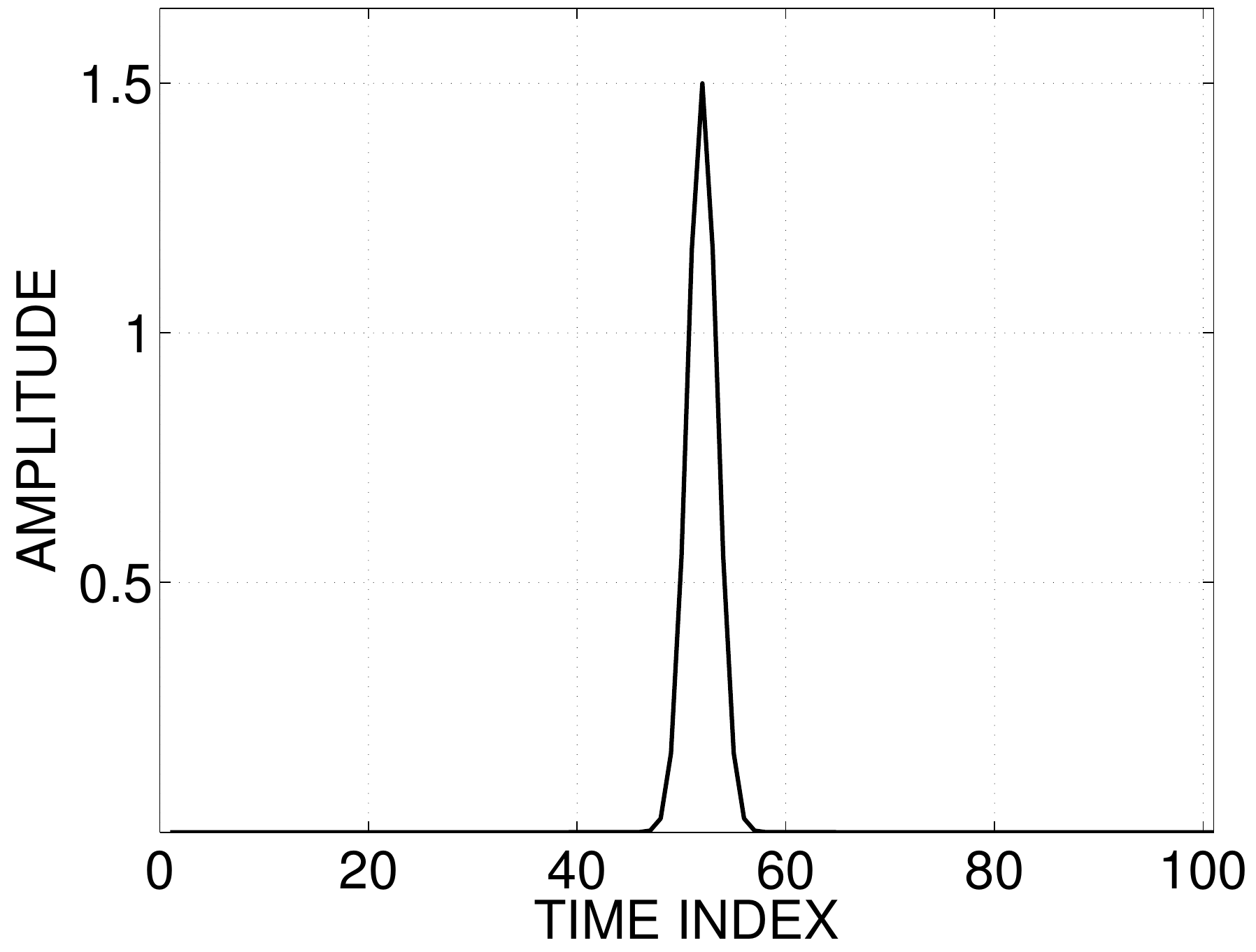}}
			\subfigure[\scriptsize{Recovery SNR versus input SNR}]{
			\includegraphics[width=1.5in]{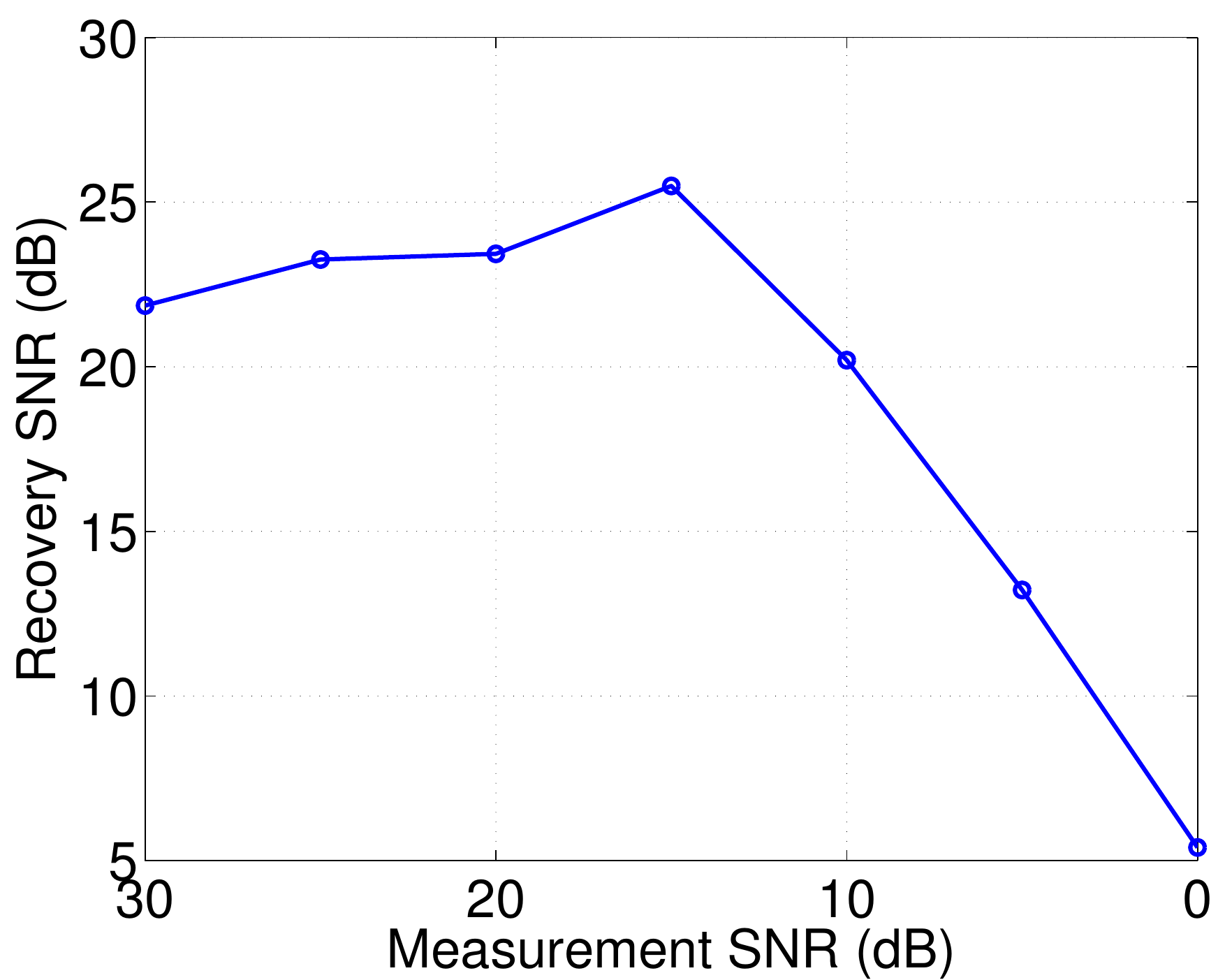}}
			\caption{\small{Performance of the BSR algorithm in the presence of noise.}}
			\vspace{-0.15in}
\label{noise_perf_fig}
\end{figure}

To conduct an empirical study of the BSR algorithm under measurement noise, we carry out experiments considering noisy measurements of the form $y_i = \text{sgn}\left( \left(AH \bold x -\tau \right)_i+w_i\right)$, where $w_i$ denotes the measurement noise. If $\mathcal{F}$ denotes the set measurements whose signs get flipped due to noise, we observe that the condition \eqref{measurement_eq1} is violated for every $i \in \mathcal{F}$, thereby rendering \eqref{noiseless_SVM_multibit_nondegen_cvx} infeasible. To circumvent this problem, we solve a relaxed version of \eqref{noiseless_SVM_multibit_nondegen_cvx}, given by   
\begin{equation}
\underset{\bold x, \tau, \xi_i}{\min}\text{\,\,}\left\| \Lambda^{(p)} \bold x \right\|_1+\beta\sum_{i}^{m}{\xi_i}  \text{\,\,s.t.\,\,}y_i \left(\phi_i^T \bold x -\tau\right)\geq 1-\xi_i,\text{\,} \xi_i\geq 0,
\label{noiseless_SVM_multibit_nondegen_cvx_relaxed}
\end{equation} 
where $\beta>0$ is an appropriately chosen constant. The measurement consistency requirement is relaxed by introducing a set of non-negative slack variables $\xi_i$.\\ 
\indent The experiment is performed on the superposition of point sources in 1-D, as shown in Fig.~\ref{OneD_illustration_figure}(a). The Gaussian blur kernel shown in Fig.~\ref{noise_perf_fig}(a) is used as the impulse response of the measurement device.~The number of binary measurements acquired is $m=2l_z=600$. The number of iterations is chosen as $I=8$. The reconstruction SNR, as a function of input/measurement SNR, is plotted in Fig. \ref{noise_perf_fig}(b). We observe that the BSR algorithm is capable of tolerating noisy measurements up to an SNR of $15$ dB. If the input SNR reduces below $15$ dB, the drop in the reconstruction SNR is gradual and acceptable. We chose $\beta=0.02$ in the experiment and found that it works reasonably well for a range of input SNR values. However, the choice of $\beta$ for optimum recovery performance requires further investigation and will be addressed separately. As a rule of thumb, one should use larger (smaller) values of $\beta$ for higher (lower) measurement SNR. 
\vspace{-0.10in}
\section{Conclusions}
We propose an iterative convex programming-based algorithm, referred to as BSR, for recovering point sources from binary measurements of the random projections of its blurred counterpart. Our algorithm does not assume the knowledge of the number of impulses to be recovered and the threshold used for binary encoding.~We have demonstrated, using 1-D and 2-D examples, that the BSR algorithm recovers the locations and the amplitudes of the point sources accurately, even under considerable blurring. The binary measurement scheme can potentially result in faster acquisition and lesser storage requirement, without compromising on the performance. We have also developed a modification of the BSR algorithm to work with noisy measurements and demonstrated that it performs reliably up to a measurement SNR of $15$ dB.

\bibliographystyle{IEEEbib}
\bibliography{strings,refs}

\vfill
\pagebreak

\bibliographystyle{IEEEbib}
\bibliography{strings,refs}

\end{document}